# ORCA: A Comprehensive AI-Driven Platform for Digital Pathology Analysis and Biomarker Discovery

Noor Shaker[1], Mohamed AbouZleikha[1] and Nuha Shaker[2]

## Abstract

Digital pathology has emerged as a transformative approach to tissue analysis, offering unprecedented opportunities for objective, quantitative assessment of histopathological features. However, the complexity of implementing artificial intelligence (AI) solutions in pathology workflows has limited widespread adoption. Here we present ORCA (Optimized Research and Clinical Analytics), a comprehensive no-code AI platform specifically designed for digital pathology applications. ORCA addresses critical barriers to AI adoption by providing an intuitive interface that enables pathologists and researchers to train, deploy, and validate custom AI models without programming expertise. The platform integrates advanced deep learning architectures with clinical workflow management, supporting applications from tissue classification and cell segmentation to spatial distribution scoring and novel biomarker discovery. We demonstrate ORCA's capabilities through validation studies across multiple cancer types, showing significant improvements in analytical speed, reproducibility, and clinical correlation compared to traditional manual assessment methods. Our results indicate that ORCA successfully democratizes access to state-of-the-art AI tools in pathology, potentially accelerating biomarker discovery and enhancing precision medicine initiatives.

## Introduction

The digitization of pathology has fundamentally transformed how tissue-based diagnostics are approached in modern medicine. Whole slide imaging (WSI) technology now enables the capture of entire histological sections at sub-micron resolution, generating datasets that contain unprecedented detail about tissue architecture, cellular morphology, and protein expression patterns [1,2]. This technological revolution has created both extraordinary opportunities and formidable challenges for the pathology community. While digital slides offer the potential for quantitative, reproducible analysis at scale, the sheer volume and complexity of data generated—often exceeding 1 gigabyte per slide—require specialized software solutions and far exceed the capacity of traditional visual assessment methods [3,4].

Parallel to these developments, a vibrant ecosystem of open-source bioimage analysis software has emerged, spearheaded by ImageJ [5] and extended by tools such as Fiji [6], Icy [7], and CellProfiler [8]. These platforms have democratized image analysis across microscopy and high-content imaging, enabling users to develop and share customized solutions in the form of plugins, pipelines, and workflows, thereby advancing reproducibility and translational research. Inspired by this model, several frameworks have been introduced for digital pathology, including OpenSlide [9], Bio-Formats [10], SlideToolkit [11], ImmunoRatio [12], Cytomine [13], QuPath

[1]SpatialX Diagnostics Inc. USA.
[2]University Pittsburgh Medical Center

[14], TIAToolbox [15], MONAI [16], SlideFlow [17], PHARAOH [18], WSInfer [19], and QuST [20]. Each of these tools contributes critical functionality—ranging from WSI format handling and tiling, to quantitative biomarker analysis, real-time visualization, and spatial multi-omics integration.

Despite these advances, the field continues to lack a unified, widely accepted software framework for digital pathology that combines accessibility for pathologists with the flexibility and performance required for state-of-the-art AI. Current solutions either require coding expertise, focus on tasks outside the scope of pathology, or only partially address the visualization and computational challenges of gigapixel WSIs. As a result, many users are forced to downsample or crop slides for analysis, apply limited general-purpose tools to subsets of their data, or revert to manual evaluation of slides, a process known to suffer from high variability and limited reproducibility [15]. At the same time, computational researchers face difficulties in deploying and distributing novel algorithms in ways that are immediately usable by practicing pathologists, slowing translation of methodological innovation into clinical impact [21].

The integration of artificial intelligence, particularly deep learning approaches, has shown remarkable promise in addressing these analytical challenges, spanning from diagnostic to prognostic applications and opening new avenues for biomarker identification, quantification, and novel biomarker discovery at a scale never possible before [1-4]. Modern AI architectures have revolutionized pathological analysis by enabling automated pattern recognition across multiple scales simultaneously—from subcellular structures to tissue-level organization patterns. Convolutional neural networks (CNNs) and transformer architectures have demonstrated superior performance in tasks ranging from cancer detection and grading to biomarker quantification and prognosis prediction [22-25]. Recent advances in vision transformers and self-supervised learning have further expanded the possibilities for discovering novel morphological patterns that may be imperceptible to human observers, potentially revealing new biomarkers with clinical significance [24].

Despite these remarkable technical advances, the translation of AI research into routine clinical practice has been hampered by several critical barriers. Chief among these is the requirement for programming expertise and deep knowledge of machine learning, which creates an insurmountable barrier for most pathologists and clinical researchers. Existing software rarely bridges this gap: research-focused tools require advanced technical skills, while commercial solutions limit flexibility and often lag behind state-of-the-art AI developments. Moreover, the absence of scalable platforms that can manage hundreds or thousands of WSIs further compounds the problem, alongside challenges in model validation, regulatory compliance, and seamless integration into existing workflows [25,26].

ORCA (Optimized Research and Clinical Analytics)[1] was specifically developed to bridge this gap by providing a comprehensive, no-code AI platform designed exclusively for digital pathology applications. Built upon over two decades of AI research and developed through close collaboration with leading pathologists and clinicians, ORCA represents a paradigm shift in how advanced computational methods are delivered to the pathology community. The platform enables users to design, train, and deploy custom AI models through an intuitive interface that requires no programming expertise, while preserving the sophistication demanded by research and clinical applications. ORCA supports the entire AI development lifecycle, including data annotation, model training and optimization, validation, and deployment in both supervised and unsupervised settings [27-34].

---

[1]Availabel at: https://spatialx-orca.com/

By abstracting complex operations behind carefully designed, intuitive interfaces, ORCA empowers pathologists and researchers to leverage their domain expertise while the platform handles the technical complexity. This democratization enables rapid hypothesis testing, accelerates biomarker discovery, and improves diagnostic reproducibility across institutions [21]. Most importantly, it lays the foundation for personalized treatment strategies by facilitating the development of predictive models tailored to specific patient populations and clinical contexts. In doing so, ORCA positions itself not merely as a software tool, but as a catalyst for transforming pathology research and clinical practice in the era of precision medicine.

## System Architecture and Design

The following sections provide detailed information about the architectural principles, technical implementation, and core functionalities that underpin ORCA's comprehensive AI platform for digital pathology. We describe the modular system design that enables seamless integration of advanced machine learning capabilities with intuitive user interfaces, the sophisticated backend infrastructure that manages large-scale pathological datasets, and the innovative workflow management systems that support the entire AI development lifecycle. Additionally, we present the platform's core analytical modules, including state-of-the-art algorithms for tissue classification, cell segmentation, spatial analysis, and biomarker quantification, along with the automated model management and deployment systems that ensure consistent performance across diverse clinical and research environments.

### Platform Overview

ORCA is architected as a cloud-native platform that seamlessly integrates data management, model development, and deployment capabilities within a unified ecosystem. The system employs a modular design philosophy, allowing users to construct custom analytical workflows by combining pre-validated components. At its foundation, ORCA incorporates state-of-the-art deep learning architectures optimized for pathological image analysis, including specialized CNNs for tissue classification, U-Net variants for segmentation tasks, and attention-based and graph-based models for spatial relationship analysis (Fig. 2).

The platform enables accessibility and seamless workflow integration through a comprehensive web-based interface accessible from any modern browser, eliminating the need for specialized software installations or hardware configurations. The user interface prioritizes accessibility and workflow efficiency through intuitive design principles that accommodate users with varying levels of technical expertise. Whole slide images can be uploaded effortlessly through drag-and-drop functionality and guided wizards that automatically handle file format validation and quality assessment. Upon upload, WSI data are automatically converted to a scalable, multi-resolution format that enables fast and smooth zooming, panning, and browsing capabilities across gigapixel images without performance degradation.

The interface integrates sophisticated annotation tools that allow users to define multiple label types, custom color schemes, and precise masks to delineate cells, tissue regions, and areas of interest with pixel-level accuracy. These annotation capabilities support both manual expert annotation and AI-assisted semi-automated approaches, enabling efficient creation of high-quality training datasets. Through this intuitive framework, users can train and validate deep learning models without writing any code, while maintaining full control over experimental design and methodology.

ORCA incorporates robust mechanisms for intelligent data partitioning and cross-validation, automatically dividing datasets into training, validation, and test sets while preserving statistical balance across classes and experimental conditions. An integrated visual interface provides comprehensive feedback on model performance, including loss curves, accuracy metrics, confusion matrices, and detailed summaries of data and annotation distributions. These tools allow users to detect potential overfitting and make informed decisions about optimization strategies. Model training can be conducted in an iterative mode: an initial model is trained on a limited set of annotated data, and its outputs are evaluated to identify performance gaps and weaknesses in the dataset. Based on this feedback, users can refine existing annotations or supply additional ones to fine-tune the model. This cycle can be repeated as needed until satisfactory performance is achieved, enabling progressive improvement through targeted annotation and retraining. Fig.1 presents a schematic overview of a standard model training cycle.

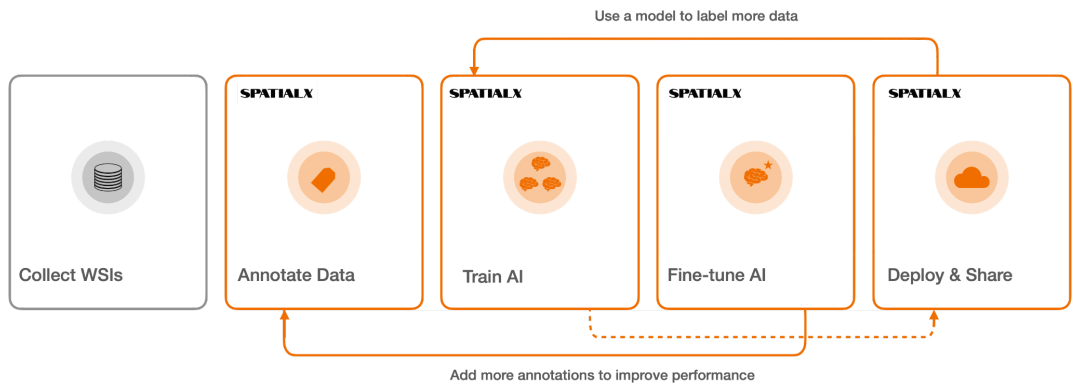

Fig. 1. Illustration of the overall workflow of building AI models using ORCA. AI models can either be trained from scratch or fine-tuned from existing model repositories. A trained model can be deployed on a subset of WSIs to visualise and inspect its performance. More data can then be added to boost performance. The cycle can be repeated a number of times until a satisfactory outcome is achieved.

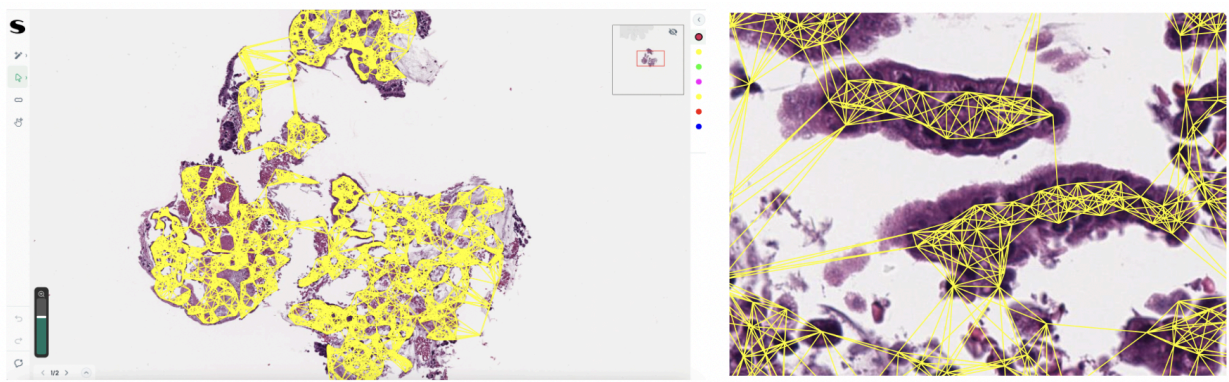

Fig 2. Illustration of ORCA's graph-based modelling applied to cells in an H&E slide. Cells are first detected and a graph is then constructed by connecting cells of a certain type within a predefined radius. Graph neural network or other forms of spatial analysis can then be applied to capture the spatial arrangement of cells.

Data processing, model training, and hyperparameter optimization are performed automatically through advanced AutoML algorithms specifically optimized for computational pathology tasks, significantly reducing the need for manual parameter tuning while ensuring optimal model performance. The platform incorporates sophisticated computational resource management that automatically scales processing power based on dataset size and model complexity, ensuring efficient utilization of cloud infrastructure. This comprehensive design philosophy ensures that ORCA remains fully accessible to pathologists and researchers regardless of their technical background, while simultaneously providing state-of-the-art AI capabilities that rival or exceed those available to computational specialists.

## Core Functionalities

### Tissue Classification and Segmentation

ORCA's tissue analysis capabilities leverage advanced deep learning architectures to provide accurate classification of tissue types, tumor regions, and cellular structures. The platform incorporates pre-trained models developed on diverse histopathological datasets, enabling rapid adaptation to new tissue types and staining protocols. Users can fine-tune these models using their own data, with the platform automatically handling data preprocessing, augmentation, and validation procedures.

### Cell Detection and Phenotyping

Cell-level analysis represents a critical capability for modern pathology applications. ORCA incorporates sophisticated cell detection algorithms that can identify and characterize individual cells within complex tissue environments. The platform supports multi-class cell classification, enabling the identification of different cell types, immune infiltrates, and pathological variants within single images and across a study cohort.

Fig 3. Illustration of ORCA's functionality. Top row- human annotations on an H&E slide and AI trained model predictions for tissue segmentation (left-to-right). Middle row- ORCA's layered tissue and cell segmentations and

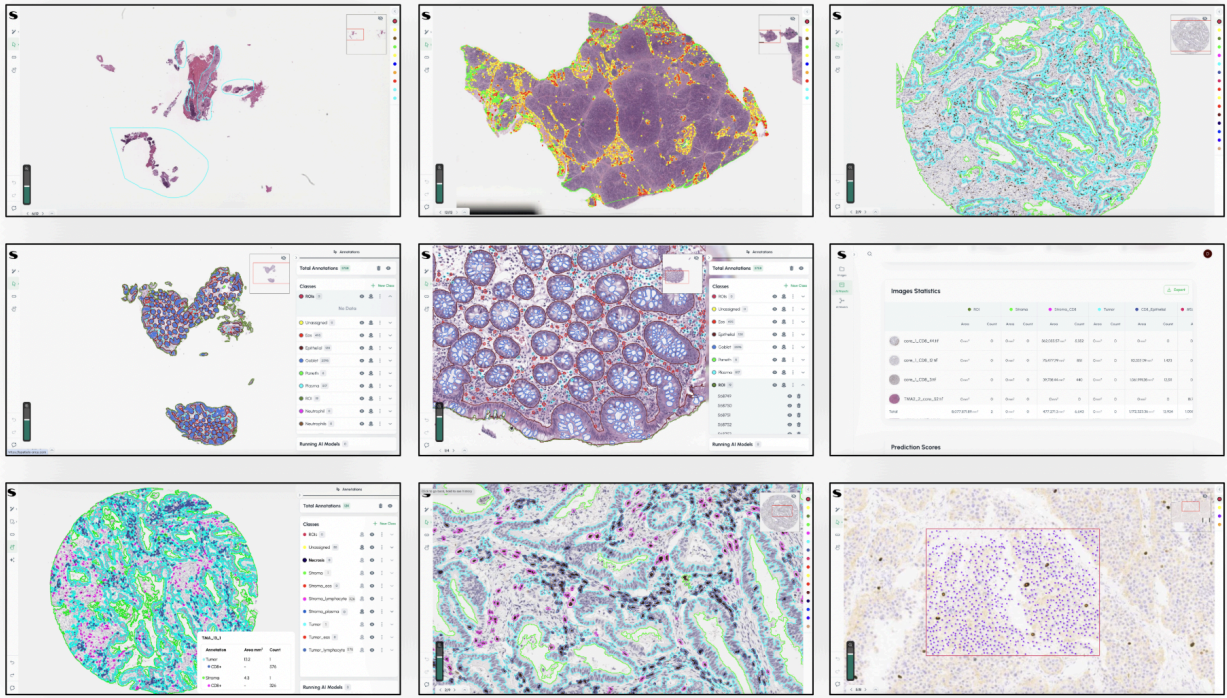

overall statistical features capturing individual slide statistics. Bottom row- cell and tissue segmentations models applied to stained images, in this case, CD8 and Ki-67 stains. The model and the visualisation tool clearly show the number and distributions of the cells within the tissues.

Morphological feature extraction algorithms quantify cellular characteristics including size, shape, borders, and spatial relationships. These features can be used to conduct deep analysis on features that correlate with clinical endpoints or can be combined with protein expression data from immunohistochemical stains to create comprehensive cellular phenotypes. ORCA's spatial analysis tools can then identify patterns of cellular organization, immune infiltration, and capture heterogeneity that may have prognostic or predictive value.

**Spatial Distribution Analysis**

Understanding the spatial organization of tissue structures is crucial for pathological interpretation and is increasingly studied as a predictive marker for treatment response [27]. ORCA provides sophisticated spatial analysis tools that can quantify the distribution of cells, identify clustering patterns, and measure distances between different tissue components. These capabilities enable the analysis of phenomena such as immune infiltration patterns, tumor-stromal interactions, and the spatial heterogeneity of biomarker expression.

The platform incorporates advanced statistical methods for spatial pattern analysis, including point process models, clustering algorithms, and network analysis approaches. Visualization tools provide intuitive representations of spatial patterns, enabling pathologists to rapidly identify regions of interest and understand complex tissue relationships.

**AI Model Management**

ORCA's model management system provides comprehensive capabilities for AI model lifecycle management. Users can access a growing repository of pre-trained models optimized for various pathological applications, with detailed performance metrics and validation studies. The platform supports model versioning, enabling users to track changes and compare performance across different model iterations.

Automated quality assessment tools evaluate model performance using established metrics and identify potential areas for improvement. The platform incorporates uncertainty quantification methods that provide confidence estimates for model predictions, enabling users to identify cases that may require additional review or alternative analytical approaches.

## Use Cases

The following section presents example use cases demonstrating how ORCA delivers state-of-the-art AI models for diverse clinical challenges, from cancer diagnosis and biomarker scoring to prognostic applications. These examples showcase both supervised (Fig. 3) and unsupervised models (Fig. 4) applied across different disease indications, illustrating ORCA's potential and clinical promise. Additional studies are documented in the literature [28-34].

### PancreaX- Diagnosis of Pancreatic Cancer

The PancreaX study [28] evaluated ORCA's ability to distinguish pancreatic ductal adenocarcinoma (PDAC) from chronic pancreatitis (CP), a notoriously difficult diagnostic challenge due to overlapping histopathological features. The dataset comprised 518 whole-slide images (WSIs) from shark core biopsies (PDAC, n = 373; CP, n = 145). A board-certified gastrointestinal pathologist used ORCA to annotate 259 WSIs by specifying benign and cancerous regions, yielding 489,556 patches (256×256 pixels) for model training and validation (70/30 split). A separate test set included 24 PDAC and 15 CP cases, and an independent cohort of 20 PDAC cases from TCGA was used for external validation.

Using ORCA, a deep learning convolutional model is trained to classify patches. The best performing model, PancreaX, achieved a sensitivity of 99% and specificity of 98% on the primary cohort. External validation on TCGA data confirmed the model's robustness, achieving 96% sensitivity without fine-tuning. These results demonstrate ORCA's ease of use, model building efficiency and capacity to generalize across institutions and datasets, providing a seamless interface for annotation and model validation and a reliable support for pathologists in challenging diagnostic contexts.

### Biliary Stricture Classification: BiliX

The BiliX study [29] investigated AI-driven diagnostic support for biliary strictures using ORCA. The dataset included 177 WSIs from 144 patients (74 adenocarcinoma cases). ORCA was used by a board-certified GI pathologist to annotate regions on the WSIs leading to 193,283 annotated patches. Data were split 70/30 for training and validation, with a separate test set of

33 WSIs (16 benign cases). Once the data is annotated, ORCA is used to train a vision transformer architecture that was then deployed in ORCA on the test set. A pathologist then investigated the outcome.

BiliX achieved a sensitivity of 93% and specificity of 79%. All tumor cases were correctly flagged by the model. These findings illustrate ORCA's seamless workflow and ease of use that allowed the pathologist, with no coding experience, to train and validate sophisticated AI models that have potential to inform clinical decision making.

### Inflammatory Bowel Disease (IBD) Histological Analysis

Histological remission is emerging as a critical treatment target in IBD, complementing endoscopic remission by providing deeper insights into disease activity and therapeutic efficacy. This study [30] aimed to develop and validate a deep learning framework comprising multiple models for the quantitative assessment of histological features in IBD. ORCA's deep learning framework was applied to quantitative histological assessment of inflammatory bowel disease. The dataset included 492 digitized H&E slides: active IBD (n = 101), chronic IBD (n = 200), and controls (n = 191). Pretrained deep learning cell segmentation models were deployed on all WSIs. New models are trained to segment tissues including epithelial regions and stroma. Cell models included goblet cells, Paneth cells, neutrophils, eosinophils, and plasma cells. Validation was performed on an independent 10% hold-out cohort. After processing all WSIs, more than 30 quantitative features were calculated that include: cell count, area, distribution, ratio, average within tissue.

The framework demonstrated strong concordance with expert annotations, producing accurate overlays for epithelial and immune features. Quantitative analysis revealed significant correlations between cell distributions and disease states. For instance, eosinophils and plasma cells were significantly elevated in active IBD compared with chronic disease ($p < 0.001$) while Goblet cell abundance was highest in controls versus both active and chronic IBD ($p < 0.001$).

These findings highlight ORCA's ability to quantify nuanced histological features, providing reproducible, objective insights into disease activity and progression beyond traditional semi-quantitative, subjective scoring.

### Ki-67 Identification and Quantification

To evaluate ORCA's capacity for automated biomarker quantification, we applied the platform to the detection and measurement of Ki-67 expression in breast cancer tissue microarrays (TMAs). The dataset included 241 cases stained for Ki-67. Manual cell counts were performed by board-certified pathologists on representative high-power fields to serve as the ground truth.

ORCA's automated pipeline accurately identified tumor nuclei and classified them by Ki-67 expression status. The model achieved a precision of 98% and recall of 99% for tumor cell detection.

These results highlight ORCA's ability to deliver reproducible, standardized biomarker quantification. By minimizing variability and reducing processing time, the platform provides a robust tool for integrating proliferative index measurements such as Ki-67 into both research and clinical workflows.

**Cancer Prognosis Analysis**

Traditional prognostic systems like the AJCC TNM staging for colorectal cancer (CRC) often fall short in predicting long-term patient outcomes. These systems often rely on limited pathologic features, leading to generalized approaches to treatment despite diverse tumor heterogeneity, such as the complex interaction between the tumor-host microenvironment. There is therefore a pressing need for more accurate, scalable tools to enhance decision-making and predict outcomes for patients with CRC. This study [31] was conducted in 191 patients with CRC, of whom a complete clinicopathologic and imaging dataset was available for 175 cases. Tumor sections were stained for CD8 to quantify tumor-infiltrating lymphocytes (TILs), with multiple intratumoral and peritumoral regions assessed to account for tumor heterogeneity. Slides were digitized and analyzed using ORCA's vision transformer-based framework. Features from quantitative assessment, such as CD8-positive TIL distributions were then integrated with clinicopathologic features including AJCC 8th edition stage, lymphovascular invasion (LVI), perineural invasion (PNI), mismatch repair (MMR) status, tumor site, metastasis status, and demographic variables to construct a prognostic model.

Patients were randomly assigned into 70% training and 30% validation cohorts. The AI model stratified patients into low- and high-risk groups based on a deep learning-derived risk score. The groups demonstrated statistically significant differences in both overall survival (OS; $p < 0.02$) and distant metastasis risk (DMR; $p < 0.0001$). For OS prediction, the most informative features included tumor site, PNI, MMR status, epithelial and stromal CD8-positive TIL densities, and five unsuervised-based vision-derived features.

These findings underscore the prognostic importance of the spatial distribution of CD8-positive TILs, which are captured quantitatively by ORCA's vision transformer architecture. By combining spatially resolved immune features with clinicopathologic variables, the model provided robust stratification beyond conventional staging systems, highlighting ORCA's potential to inform personalized treatment strategies and improve long-term patient outcomes in CRC.

**Self-Supervised Learning for Lable-Free Segmentation and Cancer Prognosis Applications**

Histopathology relies on well-established morphological features developed through centuries of clinical observation. However, human visual assessment can only capture a limited subset of the complex patterns present in tissue sections. Self-supervised deep learning models can by the model. A pathologist was then asked to interpret samples from each cluster— the descriptions are shown next to the insets. For clarity, we only highlight 7 clusters. Adopted from [32].

Fig. 4. Pathologist's interpretation of self-supervised model tissue clusters. The self-supervised model was trained to identify whether or not augmented versions of small patches of tissue came from the same original patch, without ever seeing annotations or labels. The UMAP algorithm was used to cluster and visualize the embeddings generated

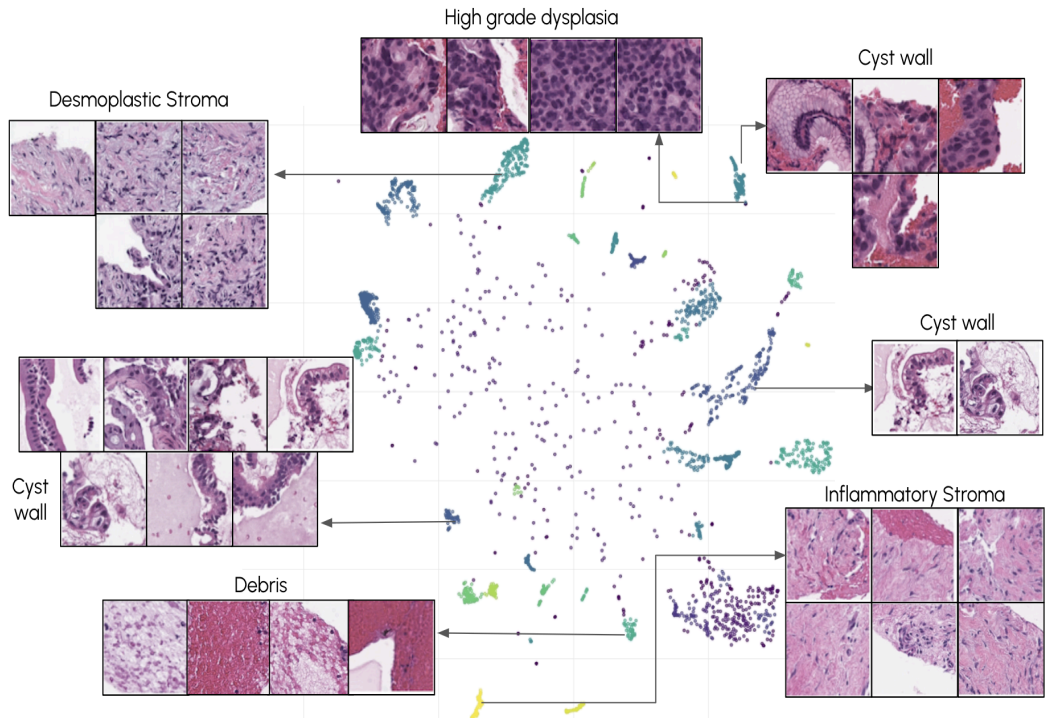

identify morphological patterns beyond conventional pathological features, potentially revealing novel tissue characteristics imperceptible to human observers. This approach opens possibilities for discovering a broader set of patterns that could enhance automated segmentation, diagnostic classification, and prognostic prediction.

We explored the use of self-supervised computer vision methods to classifying pancreatic histopathological images by learning histomorphologic features without reliance on explicit labels [32]. A dataset of 518 whole-slide images derived from shark core biopsies, including 373 PDAC and 145 CP cases, was analyzed. From these slides, 2.1 million image patches were systematically extracted using a tissue detection algorithm to ensure sampling from informative regions. A modified ResNet-50 architecture was employed to encode latent histological features in an unsupervised manner, followed by feature clustering to identify distinct tissue groups. For interpretability, a board-certified gastrointestinal pathologist examined the 20 patches closest to each of 30 cluster centroids, enabling clinical characterization of the clusters. Visualization revealed clear and distinct clusters, indicating the model's ability to capture meaningful histological features without supervision. The expert pathologist successfully identified distinct clusters corresponding to inflammatory stroma from chronic pancreatitis, desmoplastic stroma in tumor slides, cyst wall, high-grade and low-grade dysplasia, tumor tissue, and debris. Importantly, the model visualization revealed the model's ability to discard data that is similar between diseased and benign tissue and focus on distinguishing distinct tissue types.

## Discussion

ORCA represents a paradigm-shifting advancement in democratizing sophisticated AI capabilities for the global pathology community, addressing fundamental barriers that have historically limited the adoption of cutting-edge computational methods in clinical and research settings. By systematically eliminating programming requirements while maintaining state-of-the-art analytical capabilities, the platform transforms AI from a specialized technical tool accessible only to computational experts into an intuitive extension of pathological practice that expands AI applications and scales its adoption.

This democratization carries profound implications that extend far beyond mere convenience or accessibility. The platform's comprehensive approach enables pathologists to think about innovative ways to automate and improve their daily workflows, and critically, apply AI for novel marker discovery and validation studies that have the potential to improve healthcare on a wider scale. This integration has the potential to accelerate biomarker discovery exponentially by enabling rapid hypothesis testing and validation cycles that would previously have required months or years of technical development, while simultaneously improving diagnostic consistency and accuracy across diverse healthcare settings and resource environments.

Our comprehensive validation studies demonstrate that ORCA consistently achieves performance levels that meet or exceed those of expert pathologists across diverse applications while providing substantial improvements in reproducibility, efficiency, and scalability. The platform's automated approaches address long-standing challenges in pathological practice including inter-observer variability, subjective assessment limitations, and inconsistent quantification methods that have hampered multi-institutional research collaborations and clinical trial standardization. The spatial analysis capabilities represent a particularly significant advancement that opens new frontiers in pathological research and clinical practice. By enabling systematic quantification of complex spatial relationships within tissue sections, ORCA reveals biologically meaningful patterns that are difficult or impossible to assess through traditional visual examination approaches. Our validation studies demonstrate that these quantitative spatial patterns provide clinically relevant prognostic and predictive information that could fundamentally enhance treatment selection and patient stratification strategies.

ORCA's comprehensive integration of advanced AI capabilities with intuitive user interfaces distinguishes it fundamentally from existing digital pathology platforms across multiple critical dimensions. While research-focused tools like TIA Toolbox provide valuable computational pathology capabilities through sophisticated PyTorch-based frameworks, they require significant technical expertise for advanced applications and may be inaccessible to the majority of practicing pathologists and clinical researchers. Similarly, platforms like MONAI offer powerful medical AI capabilities but demand extensive programming knowledge and familiarity with deep learning concepts. Commercial platforms, although offering polished user experiences and standardized workflows, frequently suffer from limited customization capabilities that restrict their utility for novel research applications and emerging biomarker development. These systems may also provide limited access to cutting-edge AI developments due to focus on commercialization of pre-trained models and extended development cycles that lag behind rapidly evolving computational methodologies.

ORCA bridges these critical gaps by combining the analytical sophistication of specialized research tools with the accessibility and user-friendliness of commercial systems, while maintaining the flexibility needed for innovative research applications. The platform's collaborative development model, involving both leading AI researchers and practicing pathologists from premier medical institutions, ensures optimal alignment between technical capabilities and real-world clinical needs while maintaining currency with the latest computational advances. The platform's open science approach to model sharing and workflow distribution represents another significant advantage, enabling rapid dissemination of validated analytical methods across the global pathology community and fostering collaborative research initiatives that would be difficult to coordinate using traditional proprietary systems.

Despite its comprehensive capabilities and demonstrated performance advantages, ORCA has certain limitations that merit careful consideration and ongoing development attention. The platform's performance, while exceptional overall, remains dependent on the quality and representativeness of training datasets, and models may show reduced accuracy when applied to cases with highly unusual morphological features, rare pathological entities, or significant technical artifacts that fall outside established training parameters. Current validation studies have focused primarily on common cancer types and well-established biomarkers, reflecting the availability of large, well-characterized datasets for these applications. Expansion of validation studies to encompass rare diseases, emerging biomarkers, pediatric pathology applications, and diverse global population groups will be essential to ensure broad clinical utility and address potential algorithmic bias concerns that could affect equitable healthcare delivery.

The present version of ORCA concentrates primarily on histopathological applications using standard light microscopy techniques, with limited support for specialized imaging modalities including cytology, fluorescence microscopy, electron microscopy, and emerging techniques such as multiplex immunofluorescence and spatial transcriptomics. The technology underlying ORCA should be easily scalable to manage and analyze these modalities. Future development roadmaps include comprehensive expansion to these additional modalities along with integration capabilities for multi-modal data analysis combining morphological, genomic, proteomic, and clinical information.

Ongoing research initiatives focus on enhancing model robustness across diverse scanning platforms, tissue preparation protocols, and demographic populations through advanced domain adaptation techniques and federated learning approaches that can leverage distributed datasets while maintaining privacy and institutional autonomy. Additionally, active development of explainable AI capabilities will enhance model interpretability and provide pathologists with detailed insights into algorithmic decision-making processes, supporting clinical adoption and regulatory approval requirements.

ORCA's transformative capabilities position it as a catalyst for advancing precision medicine initiatives through systematic, quantitative analysis of tissue-based biomarkers that can support development of more accurate prognostic and predictive models. The platform's spatial analysis capabilities are particularly relevant for understanding complex tumor-microenvironment interactions, immune infiltration patterns, and cellular communication networks that influence treatment response and patient outcomes in ways that traditional morphological assessment cannot adequately capture. The platform's exceptional scalability and standardization capabilities make it ideally suited for large-scale population studies, multi-institutional clinical trials, and international research collaborations. Critically, unlike other computer vision interfaces, ORCA is optimized for pathology and medical applications, efficiently handling the huge size of WSIs and importantly, visualization and editing of thousands of annotations. This is of critical importance in pathology and multiplex data where images can contain thousands of cells requiring annotation and visualization, something standard interfaces and especially web-based interfaces struggle with.

## Conclusions

ORCA represents a transformative approach to AI implementation in digital pathology, successfully addressing key barriers that have limited the widespread adoption of advanced analytical methods. Through its comprehensive, user-friendly no-code platform, ORCA democratizes access to sophisticated AI capabilities while maintaining high standards of analytical quality and clinical relevance.

Our validation studies demonstrate that ORCA achieves excellent performance across diverse pathological applications, with particular strengths in reproducibility and spatial analysis capabilities. The platform's user-friendly design and workflow integration capabilities position it to accelerate biomarker discovery and enhance clinical decision-making across diverse healthcare settings.

The successful implementation of ORCA in clinical workflows demonstrates the practical value of thoughtful AI platform design that prioritizes user needs and clinical constraints. As the field of digital pathology continues to evolve, platforms like ORCA will play crucial roles in realizing the full potential of AI-driven analysis for improving patient care and advancing our understanding of disease mechanisms.

Future developments will focus on expanding the platform's capabilities to additional imaging modalities and analytical applications, while maintaining the core principles of accessibility, quality, and clinical relevance that have made ORCA a valuable tool for the pathology community.

## Acknowledgments

We thank the pathologists, clinicians, and researchers who contributed to ORCA's development and validation. Special recognition goes to the institutions that provided tissue specimens and clinical data, as well as the patients who consented to participate in these research studies.

## References

1. Hamilton, P.W. et al. Digital pathology and image analysis in tissue biomarker research. *Methods* 70, 59–73 (2014).
2. El Nahhas, O.S.M. et al. From whole-slide image to biomarker prediction: end-to-end weakly supervised deep learning in computational pathology. *Nat. Protocols* 20, 293–316 (2025).
3. Litjens, G. et al. Deep learning as a tool for increased accuracy and efficiency of histopathological diagnosis. *Sci. Rep.* 6, 26286 (2016).
4. Madabhushi, A. & Lee, G. Image analysis and machine learning in digital pathology: challenges and opportunities. *Med. Image Anal.* 33, 170–175 (2016).
5. Schneider, C.A., Rasband, W.S. & Eliceiri, K.W. NIH Image to ImageJ: 25 years of image analysis. *Nat. Methods* 9, 671–675 (2012).
6. Schindelin, J. et al. Fiji: an open-source platform for biological-image analysis. *Nat. Methods* 9, 676–682 (2012).
7. de Chaumont, F. et al. Icy: an open bioimage informatics platform for extended reproducible research. *Nat. Methods* 9, 690–696 (2012).
8. Lamprecht, M., Sabatini, D. & Carpenter, A. CellProfiler: free, versatile software for automated biological image analysis. *Biotechniques* 42, 71–75 (2007).
9. Satyanarayanan, M. et al. OpenSlide: a vendor-neutral software foundation for digital pathology. *J. Pathol. Inform.* 4, 27 (2013).
10. Linkert, M. et al. Metadata matters: access to image data in the real world. *J. Cell Biol.* 189, 777–782 (2010).
11. Nelissen, B.G.L. et al. SlideToolkit: an assistive toolset for the histological quantification of whole slide images. *PLoS One* 9, e110289 (2014).
12. Tuominen, V.J. et al. ImmunoRatio: a publicly available web application for quantitative image analysis of ER, PR, and Ki-67. *Breast Cancer Res.* 12, R56 (2010).
13. Marée, R. et al. Collaborative analysis of multi-gigapixel imaging data using Cytomine. *Bioinformatics* 32, 1395–1401 (2016).
14. Bankhead, P. et al. QuPath: Open source software for digital pathology image analysis. *Sci. Rep.* 7, 16878 (2017).
15. Pocock, J. et al. TIAToolbox as an end-to-end library for advanced tissue image analytics. *Commun. Med.* 2, 120 (2022).
16. Cardoso, M.J. et al. MONAI: an open-source framework for deep learning in healthcare. *arXiv preprint* arXiv:2211.02701 (2022).
17. Dolezal, J.M. et al. SlideFlow: deep learning for digital histopathology with real-time whole-slide visualization. *BMC Bioinformatics* 25, 134 (2024).
18. Faust, K. et al. PHARAOH: a collaborative crowdsourcing platform for phenotyping and regional analysis of histology. *Nat. Commun.* 16, 742 (2025).
19. Kaczmarzyk, J.R. et al. Open and reusable deep learning for pathology with WSInfer and QuPath. *Nat. Precis. Oncol.* 8, 9 (2024).
20. Huang, C.H., Lichtarge, S. & Fernandez, D. Integrative whole slide image and spatial transcriptomics analysis with QuST and QuPath. *npj Precis. Oncol.* 9, 70 (2025).
21. van der Laak, J., Litjens, G. & Ciompi, F. Deep learning in histopathology: the path to the clinic. *Nat. Med.* 27, 775–784 (2021).
22. Campanella, G. et al. Clinical-grade computational pathology using weakly supervised deep learning on whole slide images. *Nat. Med.* 25, 1301–1309 (2019).
23. Lu, M.Y. et al. Data-efficient and weakly supervised computational pathology on whole-slide images. *Nat. Biomed. Eng.* 5, 555–570 (2021).
24. Ciga, O., Xu, T. & Martel, A.L. Self-supervised contrastive learning for digital histopathology. *Mach. Learn. Appl.* 7, 100198 (2022).

25. Baxi, V. et al. Digital pathology and artificial intelligence in translational medicine and clinical practice. *Mod. Pathol.* 35, 23–32 (2022).
26. Niazi, M.K.K., Parwani, A.V. & Gurcan, M.N. Digital pathology and artificial intelligence. *Lancet Oncol.* 20, e253–e261 (2019).
27. Kapil, A., Spitzmüller, A., Brieu, N. et al. HER2 quantitative continuous scoring for accurate patient selection in HER2 negative trastuzumab deruxtecan treated breast cancer. Sci Rep 14, 12129 (2024).
28. Shaker, N. et al. PancreaX: Artificial intelligence–driven diagnosis of pancreatic cancer.. JCO 43, e16465-e16465(2025).
29. Shaker, N. et al. BILIX: AI-DRIVEN ENHANCEMENT FOR DIAGNOSING BILIARY STRICTURES. Gastroenterology, Volume 169, Issue 1, S-1602 (2025).
30. Shaker, N. et al. Development and Validation of a Deep Learning Framework for Quantitative Histological Analysis in Inflammatory Bowel Disease. Gastroenterology, Volume 169, Issue 1, S-1931 (2025).
31. Shaker, N. et al. A multimodal deep learning approach for the prognostic stratification of patients with colorectal cancer (CRC). JCO 43, 287-287(2025).
32. Centofanti, R. et al. Unsupervised Visual Representation Learning: Enhanced Classification of Pancreatic Tissue Types in Shark Core Biopsies. Laboratory Investigation, Volume 105, Issue 3, 102546 (2025).
33. Shaker, N. et al. Artificial Intelligence-Driven Detection of Micro and Macro Metastasis of Cutaneous Melanoma to Lymph Nodes. Laboratory Investigation, Volume 105, Issue 3, 1382 (2025).
34. Shaker, N. et al. Integrating Immune Metrics for Enhanced Colorectal Cancer Prognosis in the Era of AI. Laboratory Investigation, Volume 105, Issue 3, 682 (2025).